\documentclass[12pt]{article}
\usepackage{amssymb,amsmath}
 
\newcommand\lr[1]{{\left({#1}\right)}}

\begin{document}

\thispagestyle{empty}

\null\vskip-53pt \hfill
\begin{minipage}[t]{50mm}
HU-EP-12/20 \\
HU-MATH-2012-08\\
\end{minipage}

\vskip 3 truecm
\begin{center}

 {\Large\bf
 Three-loop universal structure constants in ${\cal N}=4$ susy Yang-Mills theory \\
 \vskip 1 cm
 or \\
 \vskip 1 cm
 Very many harmonic sums
  }
\vskip 2 truecm

{\bf  Burkhard Eden \\
}

\vskip 1 truecm

{\it Institut f\"ur Mathematik, Humboldt-Universit\"at zu Berlin, \\
Rudower Chaussee 25, Johann von Neumann-Haus, 12489 Berlin}
 \end{center}

\vskip .3truecm

\centerline{\bf Abstract}
\medskip
\noindent
We present a conjecture for the normalisation of the twist two conformal partial waves in a double OPE limit
of the four-point function of stress tensor multiplets in ${\cal N} = 4$ super Yang-Mills theory up
to three loops. This contains information about the structure constants in the OPE.

Like the twist two anomalous dimensions our result is expressed as a linear combination of
harmonic sums whose argument is the spin of the exchanged operators. 

To arrive at the result we derive asymptotic expansions for the twist two part of two unknown three-loop
integrals using the method of expansion by regions, complemented by some intuition gained on the example
of the ladder integrals up to three loops.

\newpage

\section{Introduction}

In ${\cal N} = 4$ super Yang-Mills theory
the loop corrections to the four-point function of stress-energy tensor multiplets take a factorised form:
A superconformal invariant is multiplied by an $x$-space function expressed in terms of finite conformal
integrals \cite{partialRen,hidden}. The one- and two-loop corrections to the $x$-space part have been known
for a while \cite{oneLoop,twoLoopUs,twoLoopRome}. The three-loop contribution could possibly still be
calculated using the methods of \cite{twoLoopUs} based on superconformal invariance and ${\cal N}=2$
supergraphs, but such an endeavour would surely be rather cumbersome.

However, it is not necessary to start from off-shell Feynman
graphs in order to construct the loop corrections to the four-point correlator. One can rather sort the
set of candidate scalar $l$-loop conformal integrals into orbits under an enlarged permutation symmetry
discovered in \cite{hidden}, and then fix the overall coefficient of each orbit either by appealing to the
correlator/amplitude duality \cite{EKS1to3} or by independent criteria relating to the suppression of the
highest logarithmic singularities in accordance with the expected singular behaviour of the correlator in a
Euclidean coincidence limit or in a light-cone limit, c.f. \cite{constructing}. In the latter paper the integrand of
the four-point correlator was constructed up to six loops relying on these criteria.  

In a Euclidean double OPE limit in which the positions of the four operators approach each other pairwise
we can extract anomalous dimensions from a decomposition in terms of conformal partial waves \cite{pisa}.
Here the two pairs of operators have an operator product expansion expressed as an infinite series
of other operators, and the four-point function is essentially reduced to an infinite sum over the two-point
function of these "exchanged" operators. In this paper we analyse the exchange of twist two operators.
Their anomalous dimensions are given by a universal function depending on the spin.
More precisely, perturbative calculations up to three loops have shown that this function is a linear
combination of harmonic sums \cite{osborn,klov}. This result has been of fundamental importance for the
construction of an integrable system describing the higher-loop anomalous dimensions \cite{factorised} in the
so-called $sl(2)$ or twist sector. 

The double OPE limit of the four-point function equally contains information about the structure constants in
the OPE or equivalently the three-point functions  $\langle {\cal T} \, {\cal T} \, {\cal O}^{(s)} \rangle$ where
${\cal T}$ is the stress tensor multiplet and
${\cal O}^{(s)}$ is any (twist) operator occurring in the OPE\footnote{This picture does not require an
explicit definition of the quantum corrected operators.}.
We show in this article that up to three loops the structure constants of twist two operators 
are given by a universal function written in terms of harmonic sums. Recently there has been rising interest in
an integrable systems explanation of structure constants \cite{groVieira}. We hope to foster this development
with the formulae here presented.

At one and two loops the quantum corrections to the four-point correlator are encoded in the functions
\begin{eqnarray}
F^{(1)} & = & g(1,2,3,4) \, , \\
F^{(2)} & = & \frac{1}{2} \, g(1,2,3,4)^2 \, \left( x^2_{12} x^2_{34} \, + \, 
x^2_{13} x^2_{24} \, + \, x^2_{14} x^2_{23} \right) \label{F12} \\
& + & 2 \, \left( h(1,2;3,4) \, + \, h(1,3;2,4) \, + \, h(1,4;2,3) \right)
\nonumber \, .
\end{eqnarray}
At three-loop level we use the result from \cite{hidden}
\begin{eqnarray}
&& F^{(3)} \, = \, 2 \, g(1,2,3,4) \, \left( x^2_{12} x^2_{34} \, h(1,2;3,4) + 
x^2_{13} x^2_{24} \, h(1,3;2,4) +  x^2_{14} x^2_{23} \, h(1,4;2,3) \right) \label{loop3F} \\[2mm]
&& + \, 6 \, l(1,2;3,4) + 6 \, l(1,3;2,4) + 6 \, l(1,4;2,3) + 4 \, E(1;3,4;2) + 4 \, E(1;2,4;3) + 4 \, E(1;2,3;4)
\nonumber \\[1mm]
&& + \, (x^2_{13} x^2_{24} +  x^2_{14} x^2_{23}) \, H(1,2;3,4) +
 (x^2_{12} x^2_{34} +  x^2_{14} x^2_{23}) \, H(1,3;2,4)  + \nonumber \\
 && + \, (x^2_{12} x^2_{34} +  x^2_{13} x^2_{24})  \, H(1,4;2,3) \, . \nonumber
\end{eqnarray}
This is the full answer; there are no non-planar corrections up to this order.
The definition of the integrals is
\begin{eqnarray}
g(1,2,3,4) & = & \frac{1}{4 \pi^2}
\int \frac{d^4x_5}{x_{15}^2 x_{25}^2 x_{35}^2 x_{45}^2} \, , \nonumber \\
h(1,2;3,4) & = & x^2_{34} \, \frac{1}{(4 \pi^2)^2}
\int \frac{d^4x_5 \, d^4x_6}{(x_{15}^2 x_{35}^2 x_{45}^2) x_{56}^2
(x_{26}^2 x_{36}^2 x_{46}^2)} \, , \\
l(1,2;3,4) & = & x^4_{34} \, \frac{1}{(4 \pi^2)^3}
\int \frac{d^4x_5 \, d^4x_6 \, d^4x_7}{(x_{15}^2 x_{35}^2 x_{45}^2) x_{57}^2
(x_{37}^2 x_{47}^2) x^2_{67} (x_{26}^2 x_{36}^2 x_{46}^2)} \, , \nonumber \\
E(1,2;3,4) & = & x^2_{23} x^2_{24} \, \frac{1}{(4 \pi^2)^3}
\int \frac{d^4x_5 \, d^4x_6 \, d^4x_7 \ x^2_{17}}{(x_{15}^2 x_{25}^2 x_{35}^2)
x_{57}^2 (x_{27}^2 x_{37}^2 x^2_{47}) x^2_{67} (x_{16}^2 x_{26}^2 x_{46}^2)}
\, , \nonumber \\
H(1,2;3,4) & = & x^2_{34} \, \frac{1}{(4 \pi^2)^3}
\int \frac{d^4x_5 \, d^4x_6 \, d^4x_7 \ x^2_{56}}{(x_{15}^2 x_{25}^2 x_{35}^2
x^2_{45}) x_{57}^2 (x_{37}^2 x^2_{47}) x^2_{67} (x_{16}^2 x_{26}^2 x^2_{36}
x_{46}^2)} \, . \nonumber
\end{eqnarray}
In (\ref{loop3F}) we encounter two conformal
non-ladder three-loop integrals $E,H$ (for ``easy'' and ``hard'') which are
not explicitly known. Fortunately, one can extract the twist two trajectory directly from
the integrals by the method of "expansion by regions" \cite{Smirnov}. To this end we move one point
to infinity --- which is always possible due to conformal invariance --- and then study an expansion
of the resulting three-loop integrals in terms of the remaining small distance of the OPE limit.
The procedure collapses the scalar three-point integrals to two-point tensor integrals. We evaluate
the genuine three-loop pieces by tensor reduction and an integration routine of the Mincer system
\cite{Mincer}. 
The information so obtained is not quite sufficient for our purpose so that we have to supplement it by
some intuition about the form of the results.

It is in principle possible to extend our analysis to higher twist
but we would meet more and more difficulties in extracting enough information from the integrals using
Mincer.

The article is structured as follows: In Section 1 we derive an asymptotic expansion of the box integrals
in terms of harmonic sums with only positive indices. Section 2 discusses the method of
expansion by regions and our results for the twist two part of the unknown integrals $E, H$. In Section 4
we discuss the conformal partial wave  decomposition of the correlator.

\section{Asymptotic expansion of the box-integrals}

The massless $L$-loop boxes (equivalently the $(L+1)$-rung four-point ladder
diagrams) have been evaluated in \cite{davyPhi}. They are conformal integrals
given by the appropriate weight factor times $\Phi^{(L)}(x,y)$ with the
functions
\begin{eqnarray}
\Phi^{(L)}(x,y) & = & -\frac{1}{L! (L-1)!} \, 
\int_0^1 \frac{d\xi}{y \, \xi^2 + (1-x-y) \, \xi + x} \, * \label{phidef} \\
&& \qquad \qquad \qquad \quad \; \,
 \log^{L-1}(\xi) \,
\left(\log\left(\frac{y}{x}\right) + \log(\xi) \right)^{L-1} \, 
\left(\log\left(\frac{y}{x}\right) + 2 \, \log(\xi) \right) \, .
\nonumber
\end{eqnarray}
The arguments are space time cross ratios. In a Euclidean coincidence
limit $x_{12}, \, x_{34}\, \rightarrow \, 0$ 
\begin{equation}
v \, = \, 1 - Y \, = \, \frac{x_{14}^2 x_{23}^2}
{x_{13}^2 x_{24}^2} \, \rightarrow \, 1 \, , \qquad
u \, = \, \frac{x_{12}^2 x_{34}^2} {x_{13}^2 x_{24}^2} \, \rightarrow \, 0 \, .
\end{equation}
and similar with $x_3 \leftrightarrow x_4$.
Due to the $x \leftrightarrow y$ reflection symmetry of $\Phi^{(L)}$ we may
analyse all four such situations by considering the limit $x \rightarrow 1,
\, y \rightarrow 0$ of (\ref{phidef}). Since we will be
interested in the $y^0$ part of the expansion we can conveniently put $y = 0$
in the first factor (the denominator under $d\xi$) of the integrand; the parameter integrals
remain well-defined. The simplest expansions arise for $g(1,4,2,3), \, h(1,4;2,3), \, l(1,4;2,3)$
where we put $x = 1/v, \, y = u/v$. The $\xi$-integration leads to $Li_n(-Y)$ which
is replaced by its Taylor series for small $Y$. To take the same limit on $h(1,3;2,4), \, l(1,3;2,4)$ we choose
$x = v, \, y = u$. The integration then yields $Li_n(-Y/(1-Y))$ which can also straightforwardly be expanded
in $Y$.

The point exchange $x_3 \leftrightarrow x_4$ implies the transformation
\begin{equation} \label{flipTrafo}
u \, \rightarrow \, \frac{u}{1-Y} \, , \qquad Y \, \rightarrow \,
- \frac{Y}{1-Y}
\end{equation}
on the cross ratios.
Combined with a division by $1-Y$ due to the outer weight factor this exchanges the expansions
of $h(1,4;2,3), \, l(1,4;2,3)$ in the $x_{12}, x_{34} \rightarrow 0$ limit with those of $h(1,3;2,4), \, l(1,3;2,4)$,
respectively. The one-loop box $g_{1234}$ is totally symmetric, and indeed the two choices for $x, y$ yield the
same series. 

The third case, $g(1,2,3,4), \, h(1,2;3,4), \, l(1,2;3,4)$ corresponding to the choice
\begin{equation}
x \, = \, \frac{1}{u} \, , \qquad y \, = \, \frac{v}{u} 
\end{equation}
(or its reflection) is mapped into itself. It cannot be analysed by the same simple manipulation
of the integrand in (\ref{phidef}). For the fully symmetric one-loop box this poses no problem
because the series expansion must fall upon what we had before.
In order to analyse this limit of the higher box integrals we write the first factor of the integrand as
\begin{equation}
\frac{u}{\lambda} \, \left(\frac{d\xi}{\xi - \xi_+} -
\frac{d\xi}{\xi - \xi_-} \right)\, , \quad \xi_\pm \, = \, \frac{2 - Y - u \pm
\lambda}{2 \, (1 - Y)} \, , \qquad \lambda = \sqrt{ - 4 \, u + (Y+u)^2 }
\, . 
\end{equation}
Using
\begin{equation}
\frac{1}{\xi_\pm} = 1 - x_\pm \, , \qquad x_\pm = \frac{1}{2} \left( Y + u
\pm \lambda \right)
\end{equation}
we find
\begin{eqnarray}
\frac{\lambda}{u} \, \Phi^{(2)}(Y,u) & = &
- 6 \left( Li_4(1-x_+) - Li_4(1 - x_-) \right) + 3 \, \log(1-Y) \, 
  \left( Li_3(1-x_+) - Li_3(1 - x_-) \right) \nonumber \\
&& - \frac{1}{2} \, \log^2(1-Y) \, \left( Li_2(1-x_+) - Li_2(1 - x_-) \right) 
\, , \\ \frac{\lambda}{u} \, \Phi^{(3)}(Y,u) & = &
- 20 \left( Li_6(1-x_+) - Li_6(1 - x_-) \right) + 10 \, \log(1-Y) \, 
  \left( Li_5(1-x_+) - Li_5(1 - x_-) \right) \nonumber \\
&& - 2 \, \log^2(1-Y) \, \left( Li_4(1-x_+) - Li_4(1 - x_-) \right) \nonumber \\
&& + \frac{1}{6} \, \log^3(1-Y) \, \left( Li_3(1-x_+) - Li_3(1 - x_-)
\right) \, .
\end{eqnarray}
We seek an asymptotic expansion in a Euclidean regime where $Y, u > 0$.
In the coincidence limit, $Y^2$ and $u$ are of the same order in small
quantities. Elementary trigonometry shows $- 4 \, u + Y^2 \leq 0$ to leading
order, so that the root $\lambda$ is purely imaginary. Hence $x_\pm$ are
complex with $\Re(x_\pm) > 0$. After expanding $Li_n(1-x_\pm)$ in $x_\pm$ in
this regime we treat the variables $Y$ and $u$ as independent and further
expand first in $Y$ and then in $u$. The individual polylogarithm terms
contribute $Y^n/u^m$ at order $n >= 0$, with $m = [n/2]+1/2, \,
[n/2] \, , \ldots$ up to positive powers of $u$, but negative and
half-integer powers of $u$ cancel in the complete expressions for $\Phi^{(L)}$. 
We retain only $O(u^0)$, the twist two trajectory.

When the branch point is approached from the left the polylogarithms are
described  by the following asymptotic series:
\begin{eqnarray}
Li_2(1-x) & = & \zeta_2 \, + \sum_1^\infty \frac{x^n}{n} \left(
- \left(\frac {1}{n} - \log(x) \right) \, I_0 + J_1 \right)
\, , \nonumber \\
Li_3(1-x) & = & \zeta_3 \, + \sum_1^\infty \frac{x^n}{n} \left(
- \zeta_2 \, I_o
- \left(\frac {1}{n} - \log(x) \right) \, I_1 + J_2 \right)
\, , \nonumber \\
Li_4(1-x) & = & \zeta_4 \, + \sum_1^\infty \frac{x^n}{n} \left(
- \zeta_3 \, I_o- \zeta_2 \, I_1
- \left(\frac {1}{n} - \log(x) \right) \, I_2 + J_3 \right)
\, , \label{liAsymptotics} \\
Li_5(1-x) & = & \zeta_5 \, + \sum_1^\infty \frac{x^n}{n} \left(
- \zeta_4 \, I_o - \zeta_3 \, I_1 - \zeta_2 \, I_2 
- \left(\frac {1}{n} - \log(x) \right) \, I_3 + J_4 \right)
\, , \nonumber \\
Li_6(1-x) & = & \zeta_6 \, + \sum_1^\infty \frac{x^n}{n} \left(
- \zeta_5 \, I_o - \zeta_4 \, I_1 - \zeta_3 \, I_2 - \zeta_2 \, I_3
- \left(\frac {1}{n} - \log(x) \right) \, I_4 + J_5 \right)
\nonumber
\end{eqnarray}
with the functions
\begin{eqnarray}
I_0 & = & 1 \, , \nonumber \\
I_1 & = & - \, S_1 \, , \nonumber \\
I_2 & = & -\frac{1}{2} \, S_2 + \frac{1}{2} \, S_1^2 \, , \\
I_3 & = & -\frac{1}{3} \, S_3 + \frac{1}{2} \, S_1 S_2 - \frac{1}{6} \,
S_1^3 \, , \nonumber \\
I_4 & = & -\frac{1}{4} \, S_4 + \frac{1}{3} \, S_1 S_3 + \frac{1}{8} \, S_2^2 -
\frac{1}{4} \, S_1^2 S_2 + \frac{1}{24} \, S_1^4
\nonumber
\end{eqnarray}
and
\begin{eqnarray}
J_1 & = & 0 \, , \nonumber \\
J_2 & = & S_2 \, , \nonumber \\
J_3 & = & S_3 - \, S_1 S_2 \, , \\
J_4 & = & S_4 - \, S_1 S_3 - \frac{1}{2} \, S_2^2  +
\frac{1}{2} \, S_1^2 S_2 \, , \nonumber \\
J_5 & = & S_5 - \, S_1 S_4 - \frac{5}{6} \, S_2 S_3 + \frac{1}{2} \, S_1^2 S_3 +
\frac{1}{2} \, S_1 S_2^2 - \frac{1}{6} \, S_1^3 S_2 \, . \nonumber
\end{eqnarray} 
In these formulae the range of the harmonic sums is from $1$ to $n$, so they
denote $S_1(n), S_2(n)$ etc.

The expansions are of pure transcendentality:
A sum $\sum_n x^n/n^m \, = \, Li_m(x)$ is obviously assigned weight $m$.
The more general expression $\sum_n x^n/n^m \, S_l(n)$ will be regarded as a
weight $l + m$ object if $S_l$ is any harmonic sum (or a product thereof)
of total weight $l$.
The redundant symbols $I_o$ and $J_1$ were introduced only to emphasize the
pretty iterative pattern, by which the coefficients of $\zeta$-values in the
asymptotic expansion of $Li_n(1-x)$
is known from the expansions of $Li_m(1-x)$ with $m < n$. The $I_n,J_n$
functions that we display here have been matched on Mathematica output.
Unfortunately, Mathematica runs into problems when the
order of the expansion becomes too high. The iteration as well as the fact that
$I_n,J_n$ can apparently always be expressed as products of simple
$\zeta$-values were observed on the lower examples in the list, where
the fits can contain only a very limited number of distinct structures.
Once a fit has been established it is a trivial matter to
continue the original series up to very high orders.

The asymptotic expansions for the entire box integrals (the complete
$\Phi^{(L)}$ functions including rational pre-factors) are all similar to
(\ref{liAsymptotics}). We have begun by analysing the highest
logarithms and/or $\zeta$-value contributions to understand what
type of object to fit to the series.

By the simple manipulation on the integrand of (\ref{phidef}) skeched in
the beginning of this section we found at $O(u^0)$:
\begin{eqnarray}
x^4_{13} \, g(1,4,2,3) & \rightarrow & \sum_{n=1}^\infty
 \frac{Y^{n-1}}{n} \, \left[ - \log(u)  + \frac{2}{n} \right] \, , \\
x^4_{13} \, h(1,4;2,3) & \rightarrow & \, \sum_{n=1}^\infty
 \frac{Y^{n-1}}{n} \, \biggl[ \, \frac{1}{2} \, \log^2(u) \frac{1}{n}
- \log(u) \frac{3}{n^2} + \frac{6}{n^3} \biggr] \, , \\
x^4_{13} \, l(1,4;2,3) & \rightarrow & \, \sum_{n=1}^\infty
 \frac{Y^{n-1}}{n} \, \biggl[ - \frac{1}{6} \, \log^3(u) \, \frac{1}{n^2} +
\frac{1}{2} \, \log^2(u) \,
\frac{4}{n^3} - \log(u) \, \frac{10}{n^4}
+ \frac{20}{n^5} \biggr] \, .
\end{eqnarray}
The limits of $h(1,3;2,4), \, l(1,3;2,4)$ are of a more complicated form: The resulting series can be fitted
on linear combinations of harmonic sums with exclusively positive indices:
\begin{eqnarray}
&& x^4_{13} \, h(1,3;2,4) \, \rightarrow \, \\
&& \; \sum_{n=1}^\infty \frac{Y^{n-1}}{n} \, \biggl[ \, \frac{1}{2} \,
\log^2(u) \, S_1 - \log(u) \, \left(\frac{S_1}{n} + 2 \, S_2 \right) +
\left( \frac{S_1}{n^2} + \frac{2 \, S_2}{n} - S_1 S_2 + 2 \, S_3 + 2 \, 
S_{1,2} \right) \biggr] \, , \nonumber \\ \nonumber \\
&& x^4_{13} \, l(1,3;2,4) \, \rightarrow \, \\ 
&& \; \sum_{n=1}^\infty \frac{Y^{n-1}}{n} \, \biggl[ - \frac{1}{6}
\, \log^3(u) \, \left( \frac{S_1^2}{2} + \frac{S_2}{2} \right)
+ \frac{1}{2} \, \log(u)^2 \,  \left( \frac{S_1^2}{2 \, n} +
\frac{S_2}{2 \, n} + S_1 \, S_2 + S_3 + S_{1,2} \right) \nonumber \\
&& \quad - \log(u) \, \biggl( \frac{S_1^2}{2 \, n^2} +
\frac{S_2}{2 \, n^2}
+ \frac{S_1 \, S_2}{n} + \frac{S_3}{n} + \frac{S_{1,2}}{n} + 
S_2^2 + S_1 \, S_3 + \, 2 \, S_4 + 2 \, S_{1,3} +
S_{1,1,2} - S_{1,2,1} \biggr) \nonumber \\
&& \quad + \, \frac{S_1^2}{2 \, n^3} +
\frac{S_2}{2 \, n^3}
+ \frac{S_1 \, S_2}{n^ 2} + \frac{S_3}{n^2} + \frac{S_{1,2}}{n^2}
+ \frac{S_2^2}{n} + \frac{S_1 \, S_3}{n} + \frac{2 \, S_4}{n} +
\frac{2 \, S_{1,3}}{n} + \frac{S_{1,1,2}}{n} - \frac{S_{1,2,1}}{n} \nonumber \\
&& \quad + \, 2 \, S_2 \, S_3 + S_1 \, S_4 + 3 \, S_5 + 
 3 \, S_{1,4} + S_{2,3} + 2 \, S_{1,1,3} - 2 \, S_{1,3,1} 
+ S_{2,1,2} - S_{2,2,1} - S_{1,1,2,1} + S_{1,2,1,1} \biggr] \, .\nonumber
\end{eqnarray}
In these equations --- like anywhere in this article --- all harmonic sums
have argument $n$ unless explicitly stated otherwise.

The coincidence limit on the remainig cases $h(1,2;3,4), \, l(1,2;3,4)$ requires
the more complicated procedure outlined above. We can match the results at
$O(u^0)$ by the following expressions:
\begin{eqnarray}
&& x^4_{13} \, h(1,2;3,4) \, \rightarrow \, \sum_{n=0}^\infty
 \frac{Y^{n}}{n+1} \, \biggl[ \, 6 \, \zeta(3) + S_{1,2}
- \, S_{2,1} \biggr] \, , \\
&& x^4_{13} \, l(1,2;3,4) \, \rightarrow \, \sum_{n=0}^\infty \frac{Y^{n}}
{n+1} \, \biggl[ \, 20 \, \zeta(5) + \zeta(3) \, \bigl(S_1^2 - S_2 \bigr)
\nonumber \\ && \qquad \qquad \qquad \qquad \quad
- \, S_{2,3} + S_{3,2} + \, S_{1,1,3} - S_{3,1,1}
- \, S_{1,2,2} + S_{2,2,1} - \, S_{1,1,2,1} + S_{1,2,1,1}
\biggr] \, . \nonumber
\end{eqnarray}
Although this is not manifest, these expansions --- as well as the simple
result for $g(1,4,2,3)$ --- are mapped onto themselves under the point exchange
$x_3 \leftrightarrow x_4$.

\section{Limits of $E$ and $H$ by Asymptotic Expansion}

The non-ladder integrals in the three-loop correction to the four-point function are not yet explicitly known. 
Fortunately, the method of "asymptotic expansion of Feynman integrals" \cite{Smirnov} allows us to analyse 
Feynman diagrams in any limit; coincidence limits on finite Euclidean integrals are almost the 
defining examples. Like in the case of the box integrals we will obtain the leading terms of a power series in
$Y$ at  $u^0$ and seek a fit on harmonic sums. 

Recall the definition
\begin{equation}
E(1,3;2,4) \, = \, \int \frac{d^4x_5 d^4x_6 d^4x_7 \, x_{23}^2 x_{34}^2 x_{17}^2}{x_{15}^2 x_{16}^2
x_{25}^2 x_{27}^2 x_{46}^2 x_{47}^2 x_{35}^2 x_{36}^2 x_{37}^2 x_{57}^2 x_{67}^2} \, = \,
\frac{1}{x_{13}^2 x_{24}^2} \Phi^{(E)}\left(u,v \right)
\end{equation}
where the second equality follows by conformal covariance. The integral representation is invariant under 
the exchange of points 2 and 4. As this exchanges  $u$  and $v$ we conclude that $\Phi^{(E)}(u,v) \, = \,
\Phi^{(E)}(v,u)$. Next, the exchange of points 1 and 3
has the same effect on the cross ratios, whereby we can conclude that $E(1,3;2,4) \, = \, E(3,1;2,4)$
despite of the apparent asymmetry of the integrand between $x_1$ and $x_3$. All in all we are left with
three cases to analyse: $E(1,2;3,4)$ and $E(1,3;2,4), \, E(1,4;2,3)$. In the limit $x_{12},x_{34} \rightarrow 0$,
the first case stays apart, while the other  two are related by the exchange of points 3 and 4.

\subsection{$E(1,4;2,3)$ and $E(1,3;2,4)$}

Due to conformal covariance the integrals can be uniquely reconstructed from a limit where, say, point 4 is
moved to infinity by replacing
\begin{equation} 
x_{12}^2 x_{34}^2 \, \leftrightarrow \, x_{12}^2 \, , \qquad
x_{13}^2 x_{24}^2 \, \leftrightarrow \, x_{13}^2 \, , \qquad
x_{14}^2 x_{23}^2 \, \leftrightarrow \, x_{23}^2 \, . \label{reinstate4}
\end{equation}
Note that in the limit 
\begin{equation}
u \, \rightarrow \, \frac{x_{12}^2}{x_{13}^2} \, , \qquad v \, \rightarrow \, \frac{x_{23}^2}{x_{13}^2} \, = \, 1 - 
\frac{2 (x_{12} . x_{13})}{x_{13}^2} + u \, . \label{crossLim4}
\end{equation}
We consider
\begin{equation}
E(1;2,3) \, = \, \lim_{x_4 \rightarrow \infty} \, x_4^2 \,  E(1,4;2,3) \, = \, \int \frac{d^4x_5 d^4x_6 d^4x_7 \,
x_{17}^2}{x_{15}^2 x_{16}^2 x_{25}^2 x_{27}^2 x_{36}^2 x_{37}^2 x_{57}^2 x_{67}^2} 
\end{equation}
Thanks to translation invariance we may further put $x_1 = 0$. Let us re-label $x_2 = p_1, \, x_3 = p_2,
x_5 = k_1, \, x_6 = k_2, \, x_7 = k_3$.
The integral becomes
\begin{equation}
E(p_1,p_2) \, = \,  \int \frac{d^4k_1 d^4k_2 d^4k_3 \; k_3^2}{k_1^2 k_2^2 (k_1-k_3)^2 (k_2-k_3)^2
(k_1-p_1)^2  (k_3-p_1)^2  (k_2-p_2)^2 (k_3-p_2)^2}
\end{equation}
and we are interested in the limit $p_1 \rightarrow 0$. We might simply try to expand the integrand in $p_1$ 
using
\begin{equation}
\frac{1}{(k_1 - p_1)^2} \, = \, \frac{1}{k_1^2} \sum_{n_1 = 0}^\infty \left( \frac{2 (k_1.p_1) - p_1^2}{k_1^2}
\right)^{n_1} \, , \qquad \frac{1}{(k_3 - p_1)^2} \, = \, \frac{1}{k_3^2} \sum_{n_2 = 0}^\infty \left( \frac{2 
(k_3.p_1) - p_1^2}{k_3^2} \right)^{n_2} \, . \label{expTop}
\end{equation}
These equations are only valid if $k_1^2, k_3^2 > p_1^2$, of course, and for example in calculations with
orthogonal polynomials one would indeed subdivide the integration domains according to the validity of such
expansions. Here we rather put the measure into $D \, = \, 4 - 2 \, \epsilon$ dimensions in order to regularise 
the IR singularities $1/(k_1^2)^{m_1}$ and $1/(k_3^2)^{m_2}$ that arise by extending the integration domain
to the origin. We have expanded the integrand according to the formal assigment $k_1,k_2,k_3 = O(p_2)
>> p_1$. We call this the "top region". In the "bottom region"
we declare $k_1,k_2,k_3 = O(p_1) << p_2$ and likewise employ the geometric series to expand the $k_2-
p_2$ and $k_3-p_2$ propagators in $k_2,k_3$, respectively.  Here we find UV poles arising from
the part of the integration domain where the momenta are large. The method of "expansion by regions" 
consists of evaluating not only the top and bottom regions, but the sum of all eight possibilities arising from
$k_i = O(p_1)$ or $O(p_2)$. All singularities cancel and the logarithms combine into powers of $\log{u}$.
The expansion by regions is equivalent to the "expansion by subgraphs" which in turn has been proven by
renormalisation theory to yield valid asymptotic expansions  \cite{Smirnov}. 

In massless theories, in some regions one encounters "no-scale" integrals $\int d^Dk / (k^2)^\alpha \, = \, 0$.
In the case at hand we find non-vanishing contributions in the regions
\begin{eqnarray}
& R_1 \, : \, k_1,k_2,k_3 \sim p_2 \, , \qquad \; \;
& R_2 \, : \, k_2,k_3 \sim p_2; \; k_1 \sim p_1 \, , \\
& \qquad R_3 \, : \, k_2 \sim p_2; \; k_1,k_3 \sim p_1 \, , \qquad 
& R_4 \, : \, k_1,k_2,k_3 \sim p_1 \, . \nonumber
\end{eqnarray}
Here $R_1, \, R_4$ are the "top" and "bottom" problems mentioned in the last paragraph. Quite generally, the
top problem is an $l$-loop integral with very high indices (exponents) on some propagators and therefore the
hardest to solve. The bottom problem is often of the same topology but the indices are all as in the original 
integral. The mixed cases break into an $m$-loop integral depending on the small scale and an $(l-m)$-loop
integral depending on the large scale.

The original integral $E(1;2,3)$ has dimension $[1/p^2]$. Let us consider the bottom problem $R_4$: To
lowest  order we put $k_2-p_2, k_3-p_2 \rightarrow p_2$. It follows that the corresponding three-loop integral
must go as $p_1^2$. We conclude that the bottom problem does not contribute at order $u^0$.
The same applies to $R_3$: To lowest order we have $1/(k_3 - p_2)^2 \rightarrow 1/p_2^2$. We find a
one-loop integral that has dimension of $1/(p_2^2)^{1+\epsilon}$, so once again the two-loop integral 
depending on $p_1$ must produce at least one power of $u$.

The leading term of the top problem has $p$-dependence  $1/(p_2^2)^{1+ 3 \epsilon}$, that of $R_2$ is
$1/((p_2^2)^{1+ 2 \epsilon} (p_1^2)^\epsilon)$. Since there are only two contributing regions we should find
that the leading poles from $R_1, R_2$ are equal and of opposite sign. Further, a pole $\epsilon^{-n}$ yields
logarithms to the $n$-th power in the finite part of the $\epsilon$ expansion. The logarithms from the two regions can only combine into $\log(u)$, if the leading singularity is a simple pole.

We are finally in a position to consider the top region $R_1$ in detail. We can drop $p_1^2$ from
(\ref{expTop}) because we want to restrict to $O(u^0)$. We wish to calculate
\begin{equation}
E(p_1,p_2)^{\text{top}}_{u^0} \, = \,  \sum_{n_1,n_2=0}^\infty \, \int \frac{d^Dk_1 d^Dk_2 d^Dk_3 \, \, (2
(k_1.p_1))^{n_1}  (2 (k_3.p_1))^{n_2}} {(k_1^2)^{2 +n_1} (k_1-k_3)^2  \, k_2^2  \, (k_3^2)^{n_2}
(k_2-k_3)^2 (k_2-p_2)^2 (k_3-p_2)^2} \, . \label{e1423Top}
\end{equation}
In terms of the classification of three-loop propagator type integrals in \cite{Mincer} this integral is of topology
$O_2$, the two-loop master $T_1$ with a bubble insertion in an outer line. The $k_1$ integration (i.e. the
momentum going around the bubble) can be executed using
\begin{equation}
\int \frac{d^Dk \; P_s(p)}{(k^2)^\alpha ((k-p)^2)^\beta} \, = \, \label{oneLoopInt}
\frac{1}{(p^2)^{\alpha+\beta-D/2}} \, \sum_{i=0}^{[s/2]} G(\alpha,\beta,s,i) \left\{ \frac{(k^2)^i}{i!}
\left( \frac{\square_k}{4} \right)^i\, P_s(k) \right\}_{k=p}
\end{equation}
where $P_s(k)$ is any polynomial of order $s$ and $\square_k \, = \, \partial^2 / \partial k^\mu
\partial k_\mu$. Last,
\begin{equation}
G(\alpha,\beta,s,i) \,= \, \frac{\Gamma(\alpha + \beta + i - D/2) \, \Gamma(D/2 - \alpha + s - i) \, \Gamma(D/2 - 
\beta + i)}{\Gamma(\alpha) \, \Gamma(\beta) \, \Gamma(D - \alpha - \beta+s)} \, .
\end{equation}
In our case the numerator polynomial is $k_1^{\mu_1} \ldots k_1^{\mu_{n_1}}$, all contracted onto $p_1$. In
applying (\ref{oneLoopInt}) to (\ref{e1423Top}) we discard terms with $i> 0$ since they yield $p_1^2$. It follows
\begin{equation}
E(p_1,p_2)^{\text{top}}_{u^0} \, = \,  \sum_{n_1,n_2=0}^\infty \, G(2+n_1,1,n_1,0) \, \int \frac{d^Dk_2 d^Dk_3
\, \,   (2 (k_3.p_1))^{n_1 + n_2}} {k_2^2  \, (k_3^2)^{1+\epsilon +n_1+n_2}
(k_2-k_3)^2 (k_2-p_2)^2 (k_3-p_2)^2} \, . \label{e1423TopSecond}
\end{equation}
The region $R_2$ gives a very similar result: We expand $1/(k_3-p_1)^2$ in $p_1$ and $1/(k_3-k_1)^2$ in
$k_1$. Powers of $k_1^2/k_3^2$ can be dropped because they vanish under the $k_1$ integral,
which is a one-loop bubble with ingoing momentum $p_1$. It multiplies a $T_1$ topology depending on $p_2$.
Using (\ref{oneLoopInt}) for the bubble we find
\begin{equation}
E(p_1,p_2)^{R_2}_{u^0} \, = \,  \frac{1}{(p_1^2)^\epsilon} \, \sum_{n_1,n_2=0}^\infty \, G(1,1,n_1,0) \, \int
\frac{d^Dk_2 d^Dk_3 \, \,   (2 (k_3.p_1))^{n_1 + n_2}} {k_2^2  \, (k_3^2)^{1+n_1+n_2}
(k_2-k_3)^2 (k_2-p_2)^2 (k_3-p_2)^2} \, . \label{e1423R2Second}
\end{equation}
From the last two equations it is already clear that the leading poles will cancel:
\begin{equation}
G(2+n_1,1,n_1,0)|_{\epsilon^{-1}} \, = \, \frac{1}{\epsilon} \, \frac{1}{n+1} \, = \, - \, G(1,1,n_1,0)|_{\epsilon^{-1}}
\end{equation}
Recall that the first two arguments of the $G$ function label the indices of the propagators in a one-loop
bubble integral, reflecting the fact that  $G(2+n_1,1,n_1,0)$ has a pole of IR origin while $G(1,1,n_1,0)$
contains a UV divergence. This is a very direct illustration of the cancellation of singularities due to extension
of the integration domains to the whole of Minkowski space. The remaining two loop integrals in
(\ref{e1423TopSecond}) and (\ref{e1423R2Second}) are equal to leading order in $\epsilon$. The traceless
part of the numerator (i.e. no $p_1^2$) always leads to finite integrals so that the recombination of the
logarithms can take place as we had anticipated.

Summing up, conformal invariance has permitted to reduce a four-point integral to a (generic) three-point
one. The strategy of "expansion by regions" makes it possible to extract a power series in a small ingoing
"momentum" $p_1 \, = \, x_{21}$ whereby the three-point integral collapses to a collection of two-point
problems on the expense of introducing numerators with open indices.  

Even if there are numerators, the "rule of the triangle" \cite{Mincer} can be used to solve the two-loop master 
topology $T_1$ with all integer exponents as in (\ref{e1423R2Second}) or one non-integer exponent on an
 outer line as in (\ref{e1423TopSecond}). The triangle rule is an IBP identity\footnote{integration by 
parts} which reduces the exponents of one of the $k_2, \, k_2-k_3, \, k_2-p_2$ lines in 
favour of  increasing those of $k_3, \, k_3 - p_2$. In the case at hand a single application of the triangle rule
would be sufficient because the lines involving $k_2$ all have exponent one. Hence one of them will be
cancelled in every term upon which the integral can be evaluated by iterated use of equation
(\ref{oneLoopInt}). 

In practice we rather implement a different strategy: We will use tensor reduction in order to pull $p_1$ off the
integrals and then apply the rule of the triangle to the resulting scalar integrals. In the top and bottom problems 
of the expansion by regions in other limits on $E$ and $H$ we encounter the three-loop $FA$ topology which
is implemented in the powerful "Mincer" system \cite{Mincer}. Since this programme only deals with
scalar integrals we have to understand the tensor reduction at any rate. 

Let an $l$-loop integral depending on a single outer scale $p_2$ have a numerator $N^{\mu_1 \ldots \mu_s}$
with $s$ open indices and some denominator $D$. Upon integration
\begin{equation}
\int \frac{d^Dk_1 \ldots d^Dk_l \, N^{\mu_1 \ldots \mu_s}}{D} \, = \, P_0^{\mu_1 \ldots \mu_s}(p_2) \, I_0(p_2)
+ \ldots + P_{[s/2]}^{\mu_1 \ldots \mu_s}(p_2) \label{tensorRed}
\end{equation}
where
\begin{equation}
P_i^{\mu_1 \ldots \mu_s}(p_2) \, = \, \frac{(p_2^2)^i}{i!} \left( \frac{\square_{p_2}}{4} \right)^i \, p_2^{\mu_1}
\ldots p_2^{\mu_s} \, .
\end{equation}
As in (\ref{oneLoopInt}) the d'Alembertian replaces $p_2^{\mu_i} p_2^{\mu_j}$ by $\eta^{\mu_i \mu_j}$ in a
totally symmetric fashion. Symbolic differentiation gives a fast and simple algorithmic realisation
of the tensor decomposition because we need not explicitly symmetrise.

In order to determine the $I_i$ in (\ref{tensorRed}) we project the entire equation
with $P_0 (p_2), \ldots, P_{[s/2]}(p_2)$ obtaining
\begin{equation}
J_i \, = \, M_{ij} \, I_j \, , \qquad M_{ij} \, = \, \left\{ \frac{(k^2)^i}{i!} \left(  \frac{\square_k}{4} \right)^i \;
\frac{(p_2^2)^j}{j!}\left( \frac{\square_{p_2}}{4} \right)^j \; (k . p_2)^s \right\}_{k = p_2} \, ,
\end{equation}
\begin{equation}
J_i \, = \, \left\{ \frac{(k^2)^i}{i!} \left( \frac{\square_k}{4} \right)^i \, \int \frac{d^Dk_1 \ldots d^Dk_l \;\; 
k^{\mu_1} \ldots k^{\mu_s} \, N_{\mu_1 \ldots \mu_s}}{D} \right\}_{k = p_2} \, .
\end{equation}
The matrix $M_{ij}$ depends on the regulator because $\partial_p . p \, = \,  D \, = \, 4 - 2 \,
\epsilon$. However, the tensor reduction procedure is well-defined also in exactly four dimensions, so 
that $M$ must have a power series expansion in $\epsilon$. To construct the inverse up to and including
$\epsilon^3$ we only have to invert the lowest term and complete it by linear perturbation theory; in practice 
the whole procedure was still fast up to spin 80 in the most straightforward implementation. In addition, we
can limit our scope to the construction of $I_0$ because the trace terms in 
$P_i(p_2) , \, i > 0$ obviously yield powers of $u$ upon multiplication with $p_1$. Finally, we point out that all 
steps of the algorithm can be performed on the numerator $N$ before calling any integration routine.

In our results we can finally replace $p_1, p_2$ by $x_{21}, x_{31}$, respectively, and then recover the
dependence on the fourth point by identifying $2 (x_{12} . x_{13})/x_{13}^2 \, = \, Y + O(u)$ and extending the
overall denominator to $x_{13}^2 x_{24}^2$, c.f.  (\ref{reinstate4}) and (\ref{crossLim4}).  The limit of
$E(1,3;2,4)$ can be obtained from this expansion by the transformation (\ref{flipTrafo}) corresponding to
the exchange of points 3 and 4 and division by $(1-Y)$ owing to the overall denominator. The asymptotic
series so obtained have very much the same features as the limits of the box integrals.  By a short
Mathematica script for the evaluation of the $T_1$ integrals via the triangle rule we could easily
generate the asymptotic expansion up to order $Y^{31}$, which is sufficient to determine all coefficients 
in fits on harmonic sums with only positive indices (possibly divided by powers of their arguments):
\begin{eqnarray}
&& x^4_{13} \, E(1,4;2,3) \, \rightarrow \\
&& \sum_{n=1}^\infty \, \frac{Y^{n-1}}{n} \, \biggl[ - \log(u) \, \zeta(3) \Bigl(6 \, S_1 \Bigr) - \log(u) \left(
\frac{S_1^2}{n^2} - \frac{S_1 \, S_2}{n} - S_{1, 3} + S_{3, 1} + 2 \, S_{1, 1, 2} - 2 \, S_{2, 1, 1} \right)
\nonumber \\ && \quad + \, \zeta(3) \, \left( \frac{4 \, S_1}{n} - 4 \, S_1^2 + 12 \, S_2 \right)  \nonumber \\
&& \quad + \, \frac{2 \, S_1^2}{n^3} +
\frac{2 \, S_1 \, S_2}{n^2} - \frac{2 \, S_2^2}{n} - \frac{2 \, S_1 \, S_3}{n} - 4 \, S_{1, 4} + 4 \, S_{4, 1} +
4 \, S_{1, 1, 3} - 4 \, S_{3, 1, 1} + 4 \, S_{1, 2, 2} - 4 \, S_{2, 2, 1} \biggr] \, , \nonumber \\[3mm]
&& x^4_{13} \, E(1,3;2,4) \, \rightarrow \\
&& \sum_{n=1}^\infty \, \frac{Y^{n-1}}{n} \, \biggl[ - \log(u) \, \zeta(3) \,  \left(\frac{6}{n} \right) - \, \log(u) \,
\biggl(\frac{2 \, S_1}{n^3} - \frac{S_1^2}{2 \, n^2} - \frac{S_2}{2 \, n^2} - \frac{2 \, S_3}{n}
+ \frac{S_{1,2}}{n} - \frac{S_2^2}{2} - S_1 \, S_3
\nonumber \\ && \quad -  \, \frac{3 \, S_4}{2} + 3 \, S_{1,3} - S_{1, 2, 1} + S_{2, 1, 1} \biggr)  + \, \zeta(3) \, 
\left( \frac{10}{n^2} + \frac{4 \, S_1}{n} - 2 \, S_2 \right) + \frac{8 \, S_1}{n^4}
- \frac{5 \, S_1^2}{2 \, n^3} + \frac{3 \, S_2}{2 \, n^3} \nonumber \\ && \quad 
+ \, \frac{S_1 \, S_2}{n^2} - \frac{3 \, S_3}{n^2} - \frac{S_{1, 2}}{n^2} - \frac{S_2^2}{n} + \frac{S_1 \, S_3}{n}
- \frac{8 \, S_4}{n} + \frac{4 \, S_{1, 3}}{n}   + \, \frac{S_{1, 1, 2}}{n} - \frac{3 \, S_{1, 2, 1}}{n}
+ \frac{2 \, S_{2, 1, 1}}{n} \nonumber \\[3mm] && \quad - \, 5 \, S_1 \, S_4 - 4 \, S_2 \, S_3 - 9 \, S_5
+ 13 \, S_{1, 4} + 5 \, S_{2, 3} - 2 \, S_{1, 1, 3} - 2 \, S_{1, 3, 1} + 4 \, S_{3, 1, 1}
\nonumber \\[2mm] && \quad + \, S_{2, 1, 2} - S_{2, 2, 1} + S_{1, 1, 2, 1} - S_{1, 2, 1, 1} \biggr] \, . \nonumber
\end{eqnarray}

\subsection{$H(1,2;3,4)$}

This is a single-log limit of the $H$ integral. The derivation of an asymptotic expansion for the twist two
trajectory is very similar to the single-log limit of the $E$ integral, just that here the $x_3 \leftrightarrow x_4$
exchange maps the expansion to itself. In the limit $x_4 \rightarrow \infty$ we obtain the reduced integral
\begin{equation}
H(1,2;3) \, = \, \lim_{x_4 \rightarrow \infty} \, x_4^4 \,  H(1,2;3,4) \, = \, \int \frac{d^4x_5 d^4x_6 d^4x_7 \,
x_{56}^2}{x_{15}^2 x_{16}^2 x_{25}^2 x_{26}^2  x_{35}^2 x_{36}^2 x_{37}^2 x_{57}^2 x_{67}^2} 
\end{equation}
We put $x_5 = k_1, \, x_6 = k_2, \, x_7 = k_3 , \, x_1 = 0, \, x_2 = p_1, \, x_3 = p_2$ as before and expand
using the geometric series according to which "loop momenta" are supposed to be of order $p_1$ (small) or
order $p_2$ (large). There are two contributing regions: The top problem $k_i \sim p_2$ and $k_1 \sim p_1, \,
k_2,k_3 \sim p_2$ (and symmetrically $k_1 \leftrightarrow k_2$). The bottom problem is non-vanishing but only
comes in at $O(u)$.

The top region yields an $FA$ topology, with high indices for two propagators ($p_4, p_5$ w.r.t. the definitions
of \cite{Mincer}). The second region yields a one-loop bubble times the two-loop master $T_1$ quite as
 $R_2$
in the evaluation of $E$ in the last section. The Mincer system could derive the expansion of the top problem
only up to and including $O(Y^{20})$. This may point to an installation problem, but more likely this means that
the recursion for the evaluation of the $T_1$ master with a non-integer exponent on the central line (which is
internally encountered upon applying the rule of the triangle to the $FA$ topology) is not tabulated to
sufficiently high orders.
We obtain an asymptotic series containing $\log(u) \, \zeta(3), \, \log(u), \, \zeta(3)$ and purely rational
terms. The coefficient of $\log(u) \, \zeta(3)$ must have transcendentally weight two, and from the
numerators we could in fact easily recognise  $Y^{n-1} S_1(n) / (n+1)$.
Fits of this type are quickly found also for the
non-$\zeta$ log term and the $\zeta(3)$ term without logarithm. However, the expansion up to spin 20 does
not furnish enough information to fix all 32 constants in our ansatz $S_1(n)/(n+1)^5, \ldots, S_5(n)/(n+1),
\ldots$ for the purely rational part (all sums were taken to have positive indices only). 

To make progress we invoke invariance under $x_3 \leftrightarrow x_4$ acting on the expansion 
\begin{equation}
\lim_{x_2 \rightarrow x_1} \, H(1,2;3) \, = \, \sum_{n = 1}^\infty Y^{n-1} \Bigl( \log(u) \, a_n + b_n \Bigr) 
+ O(u) \, .
\end{equation}
On the cross ratios we have the aforementioned transformation (\ref{flipTrafo}). It remains to divide by $(1-
Y)^2$, to re-expand in $Y$ and to equate with the original series. The resulting system of equations allows to
express one half of the $a_n, b_n$ in terms of the others. We substitute the ansatz
\begin{equation}
a_n \, = \, c_{5,1} \, \frac{S_4(n)}{n+1} + \ldots + c_{2,1} \, \zeta(3) \, \frac{S_1(n)}{n+1}  \, , \qquad 
b_n \, = \, c_{6,1} \, \frac{S_5(n)}{n+1} + \ldots + c_{3,1} \, \zeta(3) \, \frac{S_2(n)}{n+1} + \ldots
\end{equation}
into the first, say, 60 such equations finding seven linear relations between the 15 constants $c_{5,i}$, one for three $a_{3,i}$ and fifteen equations on the set of 32 constants $c_{6,i}$. The actual power series derived
from the integral is consistent with these conditions and now suffices to pin down the fit at $O(u^0)$:
\begin{eqnarray}
&& x^4_{13} \, H(1,2;3,4) \, \rightarrow \\
&& \sum_{n=1}^\infty \, \frac{Y^{n-1}}{n+1} \, \biggl[ - \log(u) \, \zeta(3) \Bigl( 24 \, S_1 \Bigr) - \log(u) \left(
-2 \, S_2^2 + 4 \, S_1 \, S_3+ 2 \, S_4 -  4 \, S_{1,3} + 4 \, S_{1,1,2} - 4 \, S_{1,2,1} \right) \nonumber \\
&& \quad + \, \zeta(3) \, \left( \frac{48 \, S_1}{n + 1} - 6 \, S_1^2 + 6 \, S_2 \right) 
-\frac{4 \, S_2^2}{n + 1} + \frac{8 \, S_1 \, S_3}{n + 1} +
\frac{4 \, S_4}{n + 1} - \frac{8 \, S_{1,3}}{n + 1} +
\frac{ 8 \, S_{1,1,2}}{n + 1} - \frac{8 \, S_{1,2,1}}{n + 1} \nonumber \\[2mm]
&& \quad + \, 2 \, S_2 \, S_3 + 8 \, S_1 \, S_4 + 10 \, S_5 - 
 8 \, S_{1,4} - 12 \, S_{2,3} + 10 \, S_{1,1,3}  - 8 \, S_{1,3,1} - 2 \, S_{3,1,1}  \nonumber \\
 && \quad - \,  2 \, S_{1,2,2}+ 2 \, S_{2,2,1}  - 2 \, S_{1,1,2,1} + 2 \, S_{1,2,1,1} \biggr]
\end{eqnarray}

\subsection{$E(1,2;3,4)$}

We send point 3 to infinity and identify $x_1 = p_1, \, x_2 = 0, \, x_4 = p_2, \, x_5 \, = \, k_1, \, x_6 = k_2, \,
x_7 = k_3$. The integral to expand is
\begin{equation}
E(p_1,p_2) \, = \, p_2^2  \, \int \frac{d^Dk_1 d^Dk_2 d^Dk_3 \; (k_3^2 - 2 \, k_3.p_1 + p_1^2)}{
k_1^2 \, k_2^2 \, k_3^2 (k_1 - k_3)^2 (k_2 - k_3)^2 (k_1 - p_1)^2 (k_2 - p_1)^2 (k_1 - p_2)^2 (k_3 - p_2)^2}
\end{equation}
There are six contributing regions: The top and bottom problems, two regions with two integration momenta
treated as large and one as small, and two for the opposite. One of the latter does not contribute at $O(u^0)$.
The top problem is of topology $O_2$, the bottom one of topology $FA$ and the three other regions give
(nested) bubble integrals. For the top problem and the trivial regions we need not appeal to Mincer, for the
bottom problem we used its inbuilt $FA$ routine once again. Unfortunately the programme was not able to go
beyond $Y^{17}$. To make matters worse, the $\log^3$ part of the asymptotic expansion is easy to guess and
one sees that there are series of the type $y^{n-1} \, S_2(n) / n$ and $y^{n-1} \, S_2(n) / (n+1)$, whereby
there are twice as many constants as before.

On the other hand, the expansion by regions must make sense also if the three terms of the numerator
are treated
separately: $E(p_1,p_2)$ could be a generic three-point integral, for which the strategy should be operational
independently of the fact that it is a limit of a conformal four-point integral. Interestingly we find that the
$k_3^2$ numerator is linked to the $S(n)/(n+1)$ type expansion while the other pieces cause the $S(n)/n$
part. Consequently, running the expansion by regions separately for the two parts of the numerator we can
double the number of conditions available from the integral itself.

The integral must have point exchange
symmetry $x_1 \rightarrow x_2$. This maps the $S(n)/n$ part of the fit to both pieces, but fortunately the
$S(n)/(n+1)$ piece is sent onto itself. In this sector we could solve up to five constants. Next, in all the combinations of harmonic sums that we have encountered up to now the coefficients are integer or half
integer. Using this knowledge one can simply play through all values for the remaining undetermined coefficients in a typical range and see where the dependent constants come out with denominator
1 or 2. Admittedly, this is a somewhat experimental approach, but there is very clearly only one reasonable solution:
\begin{eqnarray}
&& x^4_{13} \, E(1,2;3,4) \rightarrow \\
&& \sum_{n=1}^\infty \frac{Y^{n-1}}{n} \, \biggl[ 
- \frac{1}{6} \, \log^3(u) \, \left(\frac{2 \, S_1}{n}-S_1^2-S_2\right)
\nonumber \\ && \quad + \, \frac{1}{2} \, \log^2(u) \, \left( \frac{4 \,
S_1}{n^2} -\frac{3 \, S_1^2}{2 \, n} + \frac{3 \, S_2}{2 \, n} - S_1 \, S_2
- 3 \, S_{1,2}\right) \nonumber \\
&& \quad - \, \log(u) \, \biggl(
\frac{6 \, S_1}{n^3} - \frac{2 \, S_1^2}{n^2} + \frac{4 \, S_2}{n^2}
    -\frac{3 \, S_1 \, S_2}{n} + \frac{S_3}{n} - \frac{2 \, S_{1,2}}{n}
    -\frac{3 \, S_2^2}{2}+\frac{5 \, S_4}{2} \nonumber \\
&& \quad - \, 5 \, S_{1,3} - S_{1,1,2}
    +2 S_{1,2,1}-S_{2,1,1} \biggr) +20 \, \zeta(5) + \zeta(3) \,
\left( -\frac{2 \, S_1}{n} -S_1^2- 5 \, S_2 \right) \nonumber \\
&& \quad + \frac{8 \, S_1}{n^4} -\frac{5 \, S_1^2}{2 \, n^3} +
\frac{13 \, S_2}{2 \, n^3} -\frac{5 \, S_1 \, S_2}{n^2} +
\frac{2 \, S_3}{n^2} -\frac{S_{1,2}}{n^2} - \, \frac{2 \, S_1 \, S_3}{n}
-\frac{5 \, S_2^2}{2 \, n} + \frac{3 \, S_4}{2 \, n} -\frac{5 \, S_{1,3}}{n}
\nonumber \\ && \quad -\, \frac{2 \, S_{1,1,2}}{n} + \frac{3 \, S_{1,2,1}}{n}
-\frac{S_{2,1,1}}{n} - \, S_2 \, S_3 + 9 \, S_5 - 4 \, S_{1,4} - 4 \, S_{2,3}
\nonumber \\ && \quad
- S_{1,1,3} + 4 \, S_{1,3,1} - 3 \, S_{3,1,1} + S_{1,2,2} -2 \, S_{2,1,2} +
S_{2,2,1}+S_{1,1,2,1}-S_{1,2,1,1} \biggr]  + \nonumber
\\ && \sum_{n=1}^\infty \frac{Y^{n-1}}{n+1} \, \biggl[ - \frac{1}{6} \, \log^3 \,
\left(S_1^2+3 \, S_2\right) + \frac{1}{2} \, \log^2(u)
\left( \frac{2 \, S_1^2}{n+1}+\frac{6 \, S_2}{n+1}+4 \, S_3+4 \, S_{1,2}\right)
\nonumber \\
&& \quad - \, \log(u) \, \biggl( \frac{4 S_1^2}{(n+1)^2}+ \frac{12 \, S_2}
{(n+1)^2} +\frac{8 \, S_3}{n+1} + \frac{8 \, S_{1,2}}{n+1} 
  -3 \, S_1 \, S_3 + S_2^2 +2 \, S_4+ 8 \, S_{1,3}
\nonumber \\ && \quad + \, S_{1,1,2}-2 \, S_{1,2,1} + S_{2,1,1} \biggr)
+ \, \zeta(3) \, \left( S_1^2+3 \, S_2 \right) +
\frac{8 S_1^2}{(n+1)^3} +\frac{24 \, S_2}{(n+1)^3}
\nonumber \\ && \quad + \, \frac{16 \, S_3}{(n+1)^2}
+\frac{16 \, S_{1,2}}{(n+1)^2} - \, \frac{6 \, S_1 \, S_3}{n+1}
+\frac{2 \, S_2^2}{n+1} +\frac{4 \, S_4}{n+1} +\frac{16 \, S_{1,3}}{n+1}
\nonumber \\ && \quad + \, \frac{2 \, S_{1,1,2}}{n+1}
-\frac{4 \, S_{1,2,1}}{n+1} +\frac{2 \, S_{2,1,1}}{n+1} -
\, 6 \, S_1 \, S_4 -S_2 \, S_3 -7 \, S_5 +10 \, S_{1,4}
+4 \, S_{2,3}
\nonumber \\ && \quad 
+ \, S_{1,1,3}-4 \, S_{1,3,1}+3 \, S_{3,1,1} - \, S_{1,2,2} + 2 \, S_{2,1,2}-S_{2,2,1}
-S_{1,1,2,1}+S_{1,2,1,1} \biggr] \nonumber
\end{eqnarray}

\subsection{$H(1,4;2,3)$ and $H(1,3;2,4)$}

The $H$ integral enjoys the same type of flip symmetry as the ladder graphs and $E$. In  order to
analyse $H(1,4;2,3)$ we may thus start from $H(2,3;1,4)$ instead. We send point 4 to infinity and identify
$x_1=0, \, x_2 = p_1, \, x_3 = p_2, x_5 = k_1, \, x_6 = k_2, \, x_7 = k_3$. In momentum space notation:
\begin{equation}
H(p_1,p_2) \, = \, \int \frac{d^Dk_1 d^Dk_2 d^Dk_3 \; (k_1^2 - 2 (k_1.k_2) + k_2^2)}{k_1^2 k_2^2 k_3^2
(k_1-k_3)^2 (k_2 - k_3)^2 (k_1 - p_1)^2 (k_1 - p_2)^2 (k_2-p_1)^2 (k_2 - p_2)^2}
\end{equation}
If $k_1,k_2 \sim p_2; \, k_3 \sim p_1$ or $k_1,k_2 \sim p_1; \, k_3 \sim p_2$ we find no-scale integrals. The
contributing regions are thus the top and bottom problems --- both of topology $FA$ --- and the two
regions $R_2: \; k_1,k_3 \sim p_2; \, k_2 \sim p_1$ (and its mirror image with $k_1 \leftrightarrow k_2$), and
$R_3: \; k_1,k_3 \sim p_1; \, k_2 \sim p_2$ (and the same with $k_1 \leftrightarrow k_2$).

The leading $1/\epsilon^3$ pole must be universal to all regions for the recombination of the logarithms into
powers of $\log(u)$ to happen. We can use the trivial cases $R_2, R_3$ to try and understand the structure
of a fit. Once again, the two types of numerator terms $ A = k_1^2 + k_2^2$ and $B = - 2 (k_1.k_2)$ lead to
different structures: The A terms give a series like $Y^{n-1} \, S(n) / n^m$, the $B$ term causes $Y^{n-1} \,
S(n) / (n+1)^m$. The Mincer system is able to deal with the top problem up to and including $Y^{19}$,
thus we obtain 20 equations on the coefficients in an ansatz for each sector. It turns out that in the
logarithm terms and the $\zeta(3)$ bit of the $A$ part only $Y^{n-1} \, S(n) / n^1$ occurs, see below. Imposing
this for the rational terms at $u^0$ as well we can comfortably fix all coefficients.

With some hindsight ($S_{1,1,1,1}, S_{1,1,1,1,1}$ and high powers of $S_1$ do not occur) the ansatz for the
$B$ part can be limited to 28 coefficients, for which there are 20 equations, so what can be done?
Exchanging $x_1 \leftrightarrow x_2$ is an active transformation mapping our case to $H(1,3;2,4)$.
The usual transformation (\ref{flipTrafo}) on the cross ratios must be followed by dividing out $(1-Y)^2$ due to
the higher weight of the integral. This leads to different behaviour under the map: We find a single 
series $Y^{n-1} \, S(n)/(n+1)^m$, where for the first time the $m=0$ cases also occur. The resulting ansatz for
the rational terms is fairly large because there are many independent harmonic sums (or products thereof)
with positive indices adding up to total transcendentally weight six.

Hence a priori we cannot expect any constraint from point exchange. As a matter of fact the
simultaneous existence of the two expansions does impose a few conditions: We do not find any condition on
the coefficients for the $B$ fit for the $\log^3(u)$ terms of $H(1,4;2,3) = H(2,3;1,4)$, but one at $\log^2(u)$,
two at $\log(u)$ and three in the rational part. Hence we can restrict our ansatz for the rational terms of
the $B$ series to five
unknown constants. Supplemented by the guess that all coefficients are integer multiples of eight it was
not hard to play through the possibilities in a likely range; again we find one solution, which is presented
below.
This "diophantine" problem may seem a weak constraint, yet the difference in complexity is absolutely
striking between the preferrred solution which we display and any other random try (where one puts integer or
half-integer guesses for the independent parameters and inspects the values of the dependent parameters).
\begin{eqnarray}
&& x^4_{13} \, H(1,4;2,3) \rightarrow \\
&& \sum_{n=1}^\infty \frac{Y^{n-1}}{n} \biggl[
-\frac{1}{6} \, \log^3(u) \, \left( 2 \, S_1^2 \right)
+\frac{1}{2} \, \log^2(u) \, \left(8 \, S_1 \, S_2 \right) \nonumber \\
&& \quad -\, \log(u) \, \left( 8 \, S_{2}^2 + 16 \, S_{1} S_{3} + 4 \, S_{4} - 8 \, S_{1, 3}
+ 8 \, S_{1, 1, 2} - 8 \, S_{2, 1, 1} \right) + \zeta(3) \, \left( - 16 \, S_1^2 \right) \nonumber \\
&& \quad + \, 24 \, S_{2} \, S_{3} + 32 \, S_{1} S_{4} + 16 \, S_5 - 32 \, S_{1, 4} + 16 \, S_{1, 1, 3}-
16\, S_{3, 1, 1} + 16 \, S_{1, 2, 2}-16 \, S_{2, 2, 1} \biggr] + \qquad \qquad \nonumber
\end{eqnarray}
\begin{eqnarray}
&& \sum_{n=1}^\infty \frac{Y^{n-1}}{n+1} \biggl[
-\frac{1}{6} \, \log^3(u) \, \left( - \frac{4 \, S_1}{n+1} - 2 \, S_1^2 + 4 \, S_2 \right) \nonumber \\
&& \quad + \, \frac{1}{2} \, \log^2(u) \, \biggl(  - \frac{8 \, S_1}{(n+1)^2} - \frac{4 \, S_2}{n+1} - 8 \, S_1 \, S_2 + 12 \, S_3 \biggr)
\nonumber \\ && \quad - \,  \log(u) \, \biggl(
\frac{8 \, S_{1}^2}{(n + 1)^2} - \frac{24 \, S_{2}}{(n + 1)^2} + \frac{8 \, S_{3}}{n + 1}
- \frac{16 \, S_{1, 2}}{n + 1}- 8 \, S_{2}^2- 16 \, S_{1} \, S_{3} +  20 \, S_{4} + 8 \, S_{1, 3} \nonumber \\
&& \quad - \, 8 \, S_{1, 1, 2} + 8 \, S_{2, 1, 1}  \biggr)
+ \zeta(3) \, \left( \frac{32 \, S_1}{n+1} + 16 \, S_1^2 - 32 \, S_2 \right) \nonumber \\
&& \quad + \, \frac{64 \, S_{1}}{(n + 1)^4} + \frac{32 \, S_{1}^2}{(n + 1)^3} - \frac{64 \, S_{2}}{(n + 1)^3}
+ \frac{32 \, S_{1} \, S_{2}}{(n + 1)^2} - \frac{16 \, S_{3}}{(n + 1)^2} - \frac{64 \, S_{1, 2}}{(n + 1)^2}
- \frac{16 \, S_{2}^2}{(n + 1)} + \frac{24 \, S_{4}}{n + 1} \nonumber \\
&& \quad - \, \frac{32 \, S_{1, 3}}{n + 1} - \frac{16 \, S_{1, 1, 2}}{n + 1} + \frac{16 \, S_{2, 1, 1}}{n + 1}
- 24 \, S_{2} \, S_{3} - 32 \, S_{1} \, S_{4} + 24 \, S_{5} + 32 \, S_{1, 4} - 16 \, S_{1, 1, 3}  + 32 \, S_{3, 1, 1}  \nonumber \\
&& \quad - \, 16 \, S_{1, 2, 2} - 16 \, S_{2, 1, 2} + 16 \, S_{2, 2, 1}\biggr] \nonumber \\
&& x^4_{13} \, H(1,3;2,4) \rightarrow \\
&& \sum_{n=1}^\infty  Y^{n-1} \, \biggl[ - \frac{1}{6} \, \log^3(u) \, \left(
-\frac{4 \, S_{1}}{(n+1)^2} + \frac{S_{1}^2}{n+1} + \frac{S_{2}}{n+1}
-2 \, S_{1} \, S_{2} + 2 \, S_{3} + 2 \, S_{1, 2} \right) \nonumber \\
&& \quad + \, \frac{1}{2} \, \log^2(u) \, \left(
-\frac{8 \, S_{1}}{(n+1)^3} - \frac{4 \, S_{2}}{(n+1)^2} +
\frac{4 \, S_{1} \, S_{2}}{n+1} - 4 \, S_{1} \, S_{3} + 12 \, S_{4}
+2\, S_{1, 2, 1} - 2 \, S_{2, 1, 1} \right) \nonumber \\
&& \quad - \, \log(u) \, \biggl( -\frac{4 \, S_{1}^2}{(n+1)^3} - \frac{12 \, S_{2}}{(n+1)^3}
+ \frac{8 \, S_{1} \, S_{2}}{(n+1)^2}- \frac{8 \, S_{3}}{(n+1)^2}- \frac{8 \, S_{1, 2}}{(n+1)^2}
+ \frac{2 \, S_{2}^2}{n+1}+ \frac{14 \, S_{1} \, S_{3}}{n+1} \nonumber \\
&& \quad - \, \frac{12 \, S_{1, 3}}{n+1} + \frac{2 \, S_{1, 1, 2}}{n+1}- \frac{2 \, S_{2, 1, 1}}{n+1}
+ 4 \, S_{2} \, S_{3} - 2 \, S_{1} \, S_{4} + 42 \, S_{5} - 14 \, S_{1, 4} - 10 \, S_{2, 3} \nonumber \\
&& \quad + \, 4 \, S_{1, 1, 3} + 4 \, S_{1, 3, 1}  - 8 \, S_{3, 1, 1}
- 2 \, S_{2, 1, 2} + 2 \, S_{2, 2, 1}-  2 \, S_{1, 1, 2, 1} + 2 \, S_{1, 2, 1, 1} \biggr) \nonumber \\
&& \quad + \, \zeta(3) \, \left( \frac{32 \, S_{1}}{(n+1)^2} - \frac{8 \, S_{1}^2}{n+1}
- \frac{8 \, S_{2}}{n+1} + 16 \, S_{1} \, S_{2} - 16 \, S_{3} - 16 \, S_{1, 2} \right) \nonumber \\
&& \quad + \, \frac{64 \, S_1}{(n+1)^5} - \frac{16 \, S_1^2}{(n+1)^4} - \frac{16 \, S_2}{(n+1)^4}
+ \frac{16 \, S_1 \, S_2}{(n+1)^3} - \frac{32 \, S_3}{(n+1)^3} - \frac{32 \, S_{1,2}}{(n+1)^3}
+ \frac{44 \, S_1 \, S_3}{(n+1)^2} - \frac{12 \, S_4}{(n+1)^2} \nonumber \\
&& \quad - \, \frac{56 \, S_{1,3}}{(n+1)^2}
+ \frac{4 \, S_{1,1,2}}{(n+1)^2} - \frac{4 \, S_{2,1,1}}{(n+1)^2}
+ \frac{16 \, S_2 \, S_3}{n+1} + \frac{44 \, S_1 \, S_4}{n+1} + \frac{20 \, S_5}{n+1}
- \frac{52 \, S_{1,4}}{n+1}- \frac{28 \, S_{2,3}}{n+1} \nonumber \\
&& \quad + \, \frac{8 \, S_{1,1,3}}{n+1}
- \frac{8 \, S_{3,1,1}}{n+1}- \frac{4 \, S_{2,1,2}}{n+1} + \frac{4 \, S_{2,2,1}}{n+1}
+ \frac{2}{3} \, S_{1}^2 \, S_{2}^2 - \frac{4}{3} \, S_{3}^2 + 22 \, S_{2} \, S_{4}
+ \frac{20}{3} \, S_{1} \, S_{5}  \nonumber \\
&& \quad + \, 108 \, S_{6} - 48 \, S_{1, 5}
- 56 \, S_{2, 4} + \frac{52}{3} \, S_{1, 1, 4}  + \frac{10}{3} \, S_{1, 4, 1} - \frac{62}{3} \, S_{4, 1, 1}  \nonumber \\
&& \quad + \, \frac{26}{3} \, S_{1, 2, 3} - \frac{4}{3} \, S_{1, 3, 2} + \frac{26}{3} \, S_{2, 1, 3}
+ \frac{38}{3} \, S_{2, 3, 1} - \frac{28}{3} \, S_{3, 1, 2} + \frac{14}{3} \, S_{3, 2, 1}  \nonumber \\
&& \quad - \, 4 \, S_{1, 1, 1, 3} - 4 \, S_{1, 1, 3, 1} + 8 \, S_{1, 3, 1, 1}  - \frac{8}{3} \, S_{1, 1, 2, 2} - \frac{2}{3} \, S_{1, 2, 1, 2}
- \frac{14}{3} \, \, S_{1, 2, 2, 1} - \frac{2}{3} S_{2, 1, 1, 2}  \nonumber \\
&& \quad - \, \frac{26}{3} \, S_{2, 1, 2, 1}
+ \frac{4}{3} \, S_{2, 2, 1, 1} + 2 \, S_{1, 1, 1, 2, 1} - 2 \, S_{1, 1, 2, 1, 1}  \biggr] \nonumber
\end{eqnarray}

\section{CPWA analysis}

At tree level, the correlation function is given by products of free scalar propagators:
\begin{align}\label{eq:1}
G^{(0)}_4(1,2,3,4) 
& =   \frac{(N_c^2-1)^2}{4 \, (4 \pi^2)^4}
\bigg[\lr{\frac{y^2_{12}}{x^2_{12}} \frac{y^2_{34}}{x^2_{34}}}^2+
 \lr{\frac{y^2_{13}}{x^2_{13}} \frac{y^2_{24}}{x^2_{24}}}^2 +
 \lr{\frac{y^2_{41}}{x^2_{41}} \frac{y^2_{23}}{x^2_{23}}}^2 \bigg] \\
& +   \frac{N_c^2 -1}{(4 \pi^2)^4} \,\
\left( \frac{y_{12}^2}{x_{12}^2} \frac{y_{23}^2}{x_{23}^2}  
\frac{y_{34}^2}{x_{34}^2} \frac{y_{41}^2}{x_{41}^2} +
 \frac{y_{12}^2}{x_{12}^2} \frac{y_{24}^2}{x_{24}^2}  
\frac{y_{34}^2}{x_{34}^2} \frac{y_{13}^2}{x_{13}^2} +
 \frac{y_{13}^2}{x_{13}^2} \frac{y_{23}^2}{x_{23}^2}  
 \frac{y_{24}^2}{x_{24}^2} \frac{y_{41}^2}{x_{41}^2} \right), \nonumber
\end{align}
The loop-corrections to $G_4$ take a factorised form \cite{partialRen,hidden}:
\begin{align}\label{intriLoops}
  G_{4}^{(l)}(1,2,3,4)= \frac{2 \, (N_c^2-1)}{(4\pi^2)^{4}} \times R(1,2,3,4)   \times  F^{(l)}(x_i) \qquad \mbox{(for $l \ge 1$)} 
\end{align} 
Here $R(1,2,3,4)$ is a universal, $l-$independent
rational function of the space-time, $x_i$, and harmonic, $y_i$, coordinates at the four external points $1,2,3,4$:
 \begin{align}\label{eq:7}
 R(1,2,3,4) &= \frac{y^2_{12}y^2_{23}y^2_{34}y^2_{14}}{x^2_{12}x^2_{23}x^2_{34} x^2_{14}}(x_{13}^2 x_{24}^2-x^2_{12} x^2_{34}-x^2_{14} x^2_{23})\notag
 \\ &
+\frac{ y^2_{12}y^2_{13}y^2_{24}y^2_{34}}{x^2_{12}x^2_{13}x^2_{24} x^2_{34}}(x^2_{14} x^2_{23}-x^2_{12} x^2_{34}-x_{13}^2 x_{24}^2)  \notag
\\
& +\frac{y^2_{13}y^2_{14}y^2_{23}y^2_{24}}{x^2_{13}x^2_{14}x^2_{23} x^2_{24}}(x^2_{12} x^2_{34}-x^2_{14} x^2_{23}-x_{13}^2 x_{24}^2) \notag 
 \\ &
+
\frac{y^4_{12} y^4_{34}}{x^2_{12}x^2_{34}}   + \frac{y^4_{13} y^4_{24}}{x^2_{13}
x^2_{24}} + \frac{y^4_{14} y^4_{23}}{x^2_{14}x^2_{23}} \,,
\end{align}
while $F^{(\ell)}(x_i)$ are functions of $x_i$ only, which are explicitly stated up to three loops in terms of the box integrals and $E$ and $H$ in formulae (\ref{F12}), (\ref{loop3F}) in the introduction.

In the OPE limit $x_2 \rightarrow x_1, \, x_4 \rightarrow x_3$ the weight 2 operators at points 1,2 and 3,4
respectively fuse into an expansion in terms of operators "exchanged" between the two halves of the
four-point function. The exchanged operators carry twist (dilatation weight - spin), spin and 
$SU(4)$ quantum numbers. Since we are fusing two operators in the \textbf{20'} representation, the
exchanged operators must carry one of the representations in the product $\mathbf{20'}\times\mathbf{20'} =
\mathbf{1} + \mathbf{15} + \mathbf{20'} + \mathbf{84} + \mathbf{105} +\mathbf{175}$.
The correlator as written in the last two formulae has six "channels" distinguished by the $y$ variables
pertaining to the internal symmetry group. It has been worked out in \cite{glebOPE} which linear combination
of these channels correspond to the exchange of operators in a given representation. In particular, if
we label the channels according to
\begin{equation}
G_4 \, = \, y_{12}^4 y_{34}^4 \, A_1 + y_{13}^4 y_{24}^4 \, A_2 + y_{14}^4 y_{23}^4 \, A_3 +
 y_{12}^2 y_{34}^2 \, y_{13}^2 y_{24}^2 \, A_4 + y_{12}^2 y_{34}^2 \, y_{14}^2 y_{23}^2 \, A_5 +
 y_{13}^2 y_{24}^2 \, y_{14}^2 y_{23}^2 \, A_6 
\end{equation}
then the \textbf{20'} exchange corresponds to $A_2 + A_3 + \frac{5}{3} \, A_4 + \frac{5}{3} \, A_5 +
\frac{1}{6} \, A_6$. Both at tree- and at loop-level we find that the leading power singularity is $1/(x_{12}^2
x_{34}^2)$ coming from $A_4, A_5$:
\begin{eqnarray}
\lim_{x_{12}, x_{34} \rightarrow 0} 
G_4^\mathbf{20'} &= & \frac{5 (N^2 -1)}{3 \, (4 \pi^2)^4 \, x_{12}^2 x_{34}^2 x_{13}^2 x_{24}^2} * \\
&& \left( \left(\frac{1}{1-Y} + 1 \right) +  2 \left( \frac{1}{1-Y} - 1 - Y \right) \sum_{l = 1}^\infty \lim_{x_{12}, x_{34}
\rightarrow 0} \, a^l \, F^{(l)}(x_i) \right)  + O(u)  \nonumber \label{corLim}
\end{eqnarray}
where $a$ is the effective coupling.
Powers of $u$ can be discarded because they correspond to higher twist. On the other hand, the expansion in
the $Y$ variable is associated to the spin of the exchanged twist two operators. 

In terms of the elementary field the operators we discuss are schematically realised as ${\cal O}^{(s)} \, = \, 
\mathrm{tr}(W D^{\{\mu_1} \ldots D^{\mu_s\}} W)$. The positioning of the trace-free symmetrised Yang-Mills
covariantised derivatives $D^\mu \, = \, \partial^\mu + i \, g \, [A^\mu, \bullet]$ on the two $W$ fields decides
about the anomalous dimension of the operators.
All these composites have twist two because there are two scalar elementary fields of dimension one carrying
the derivatives.
The spin 0 operator in this class is the protected primary ${\cal O}$.
At spin one we only find $\partial^\mu \, {\cal O}$, a "conformal
descendent" of ${\cal O}$. By tree-level orthogonalisation one sees that there is one new primary
operator at every even spin, all other combinations of derivatives give descendents of operators with lower 
spin. The descendents do show in the OPE, but their occurrence is statically linked to that of the primary field.
One can resum the contribution of the descendents into a "conformal partial wave" labelled by the primary
operator, see \cite{pisa,glebOPE,osborn0} and references therein:
\begin{equation}
\text{cpwa}(s) \, = \, N(s) \; \; u^{\gamma/2} \; Y^s \; 
_2F_1 \Bigl(s+1+ \gamma/2, s + 1 + \gamma/2, 2 + 2 \, s + \gamma; \, Y \Bigr)
\end{equation}
When expanding in the effective coupling $a$ the cpwa furnish logarithms to be matched on those of
the $F^{(l)}$ in (\ref{corLim}). Our task is thus to solve
\begin{equation}
\left( \left(\frac{1}{1-Y} + 1 \right) +  2 \left( \frac{1}{1-Y} - 1 - Y \right) \sum_{l = 1}^\infty \lim_{x_{12}, x_{34}
\rightarrow 0} \, a^l \, F^{(l)}(x_i) \right) \, = \, \sum_{s=0}^\infty \, \text{cpwa}(s)
\end{equation}
for $\gamma(s), N(s)$. As it should, the normalisation of the cpwa turns out to be zero if the spin is odd.
For even spin we find 
\begin{eqnarray}
N(s) & = &  2 \, \left( \frac{\Gamma \left(s + 1 + \frac{\gamma}{2} \right)^2}{\Gamma \left(2 \, s + 1 + \gamma
\right)} - \frac{1}{4} \, \sum_{i=2}^\infty \zeta(i) \, b_i \right)  \label{result} \\
&& * \biggl( 1 + a \, c_{1,2} + a^2 \, \Bigl( \zeta(3) \, c_{2,1} + c_{2,4}
\Bigr) +  a^3 \Bigl( \zeta(5) \, c_{3,1} + \zeta(3) \, c_{3,3} + c_{3,6} \Bigr) + \ldots \biggr) \nonumber
\end{eqnarray} 
where
\begin{equation}
a \, = \, \frac{g^2 \, N}{4 \, \pi^2} \, , \qquad
b_2 \, = \, - \, \gamma^2 + \Bigl(S_1(2 \, s) - S_1(s) \Bigr) \, \gamma^3 + \ldots \, , \qquad b_3 \, = \,
\gamma^3 + \ldots \, .
\end{equation}
The $b_i$ cancel the explicit dependence of the $\Gamma$ ratio on $\zeta$ values; the first factor
in (\ref{result}) is meant to be purely rational. The anomalous dimension
depends on the spin and has an expansion $\gamma \, = \, a \, \gamma_1 + a^2 \, \gamma_2 + a^3
\, \gamma_3 + \ldots$
\begin{eqnarray}
\gamma_1 & = &  2 \, S_1 \, , \\[2mm]
\gamma_2 & = & - \, S_{-3} - 2 \, S_{-2} \, S_{1} - 2 \, S_{1} \, S_{2} - S_{3} + 
 2 \, S_{-2, 1} \, , \\[2mm]
 \gamma_3 & = & 3 \, S_{-5} + 8 \, S_{-4} \, S_{1} + 
 S_{-2}^2 \, S_{1} + 6 \, S_{-3} \, S_{1}^2 + 
 S_{-3} \, S_{2} + 4 \, S_{-2} \, S_{1} \, S_{2} + 
 2 \, S_{1} \, S_{2}^2  \\[2mm]
 && + \, 2 \, S_{-2} \, S_{3} + 
 2 \, S_{1}^2 \, S_{3} + S_{2} \, S_{3} + 
 3 \, S_{1} \, S_{4} + S_{5} - 6 \, S_{-4, 1} - 
 12 \, S_{1} \, S_{-3, 1} - 6 \, S_{-3, 2} \nonumber \\[2mm]
 && - \, 4 \, S_{1}^2 \, S_{-2, 1} - 2 \, S_{2} \, S_{-2, 1} - 
 10 \, S_{1} \, S_{-2, 2} - 6 \, S_{-2, 3} + 
 12 \, S_{-3, 1, 1} + 16 \, S_{1} \, S_{-2, 1, 1} \nonumber \\[2mm]
 && + \,  12 \, S_{-2, 1, 2} + 12 \, S_{-2, 2, 1} - 24 \, S_{-2, 1, 1, 1} \, . \nonumber
\end{eqnarray}
In these formulae all harmonic sums depend on the argument $s$. Our result is in complete agreement with
 the literature \cite{osborn,klov}.
 
The tree-level normalisation is
easily recognised to be the ratio $2 \, (s!)^2/(2 s)!$ which we take out of the entire normalisation factor.
A fit of the one-loop normalisation on harmonic sums is then in fact possible, but only if in addition $S(2 s)$
are taken into account. The result for the coefficients is compatible with promoting the
the factorials at tree to the $\Gamma$ functions in (\ref{result}); this mimics the $Y$ dependent part of the
cpwa. At two and three loops the $\Gamma$ functions correctly incorporate all $S(2 s)$ terms.
Below we state our results for the coefficients in the second line of (\ref{result}). In these formulae all
harmonic sums have argument $s$ once again.
\begin{eqnarray}
c_{1,2} & = & - \, S_2 \, , \\[2mm]
c_{2,1} & = & 3 \, S_1 \, , \\[1mm]
c_{2,4} & = & \frac{5}{2} \, S_{-4} + S_{-2}^2 + 2 \, S_{-3} \, S_{1} + 
 S_{-2} \, S_{2} + S_{2}^2 + 2 \, S_{1} \, S_{3} + 
 \frac{5}{2} \, S_{4} - 2 \, S_{-3, 1} - S_{-2, 2} - 2 \, S_{1, 3} \, , \\
c_{3,1} & = & \, - \frac{25}{2} \, S_1 \, , \\
c_{3,3} & = & - \, 3 \, S_{-3} - 10 \, S_{-2} \, S_{1} + \frac{4}{3} \, S_{1}^3 - 
 6 \, S_{1} \, S_{2} - \frac{4}{3} \, S_{3} + 6 \, S_{-2, 1} \, , \\
c_{3,6} & = &  - \,11 \, S_{-6} + \frac{5}{2} \, S_{-3}^2 - 5 \, S_{-4} \, S_{-2} - 
 \frac{41}{2} \, S_{-5} \, S_{1} - S_{-3} \, S_{-2} \, S_{1} - 
 5 \, S_{-4} \, S_{1}^2 - 2 \, S_{-2}^2 \, S_{1}^2 \\ 
 && + \, \frac{4}{3} \, S_{-3} \, S_{1}^3 - \frac{13}{2} \, S_{-4} \, S_{2} - 
 \frac{3}{2} \, S_{-2}^2 \, S_{2} - 10 \, S_{-3} \, S_{1} \, S_{2} - 
 2 \, S_{-2} \, S_{2}^2 - S_{2}^3 - \frac{16}{3} \, S_{-3} \, S_{3} \nonumber \\
 && - \, 8 \, S_{-2} \, S_{1} \, S_{3} - 6 \, S_{1} \, S_{2} \, S_{3} - 
 3 \, S_{3}^2 - 3 \, S_{-2} \, S_{4} + 9 \, S_{1}^2 \, S_{4} - 
 4 \, S_{2} \, S_{4} + \frac{15}{2} \, S_{1} \, S_{5} - \frac{13}{2} \, S_{6} \nonumber \\
 && + \, 14 \, S_{-5, 1} + 11 \, S_{1} \, S_{-4, 1} + 9 \, S_{-4, 2} - 
 12 \, S_{1} \, S_{-3, -2} + 10 \, S_{-2} \, S_{-3, 1} - 4 \, S_{1}^2 \, S_{-3, 1} \nonumber \\
 && + 8 \, S_{2} \, S_{-3, 1} + 
 4 \, S_{1} \, S_{-3, 2} + 9 \, S_{-3, 3} - 
 10 \, S_{-3} \, S_{-2, 1} + 
 14 \, S_{-2} \, S_{1} \, S_{-2, 1} - 
 \frac{8}{3} \, S_{1}^3 \, S_{-2, 1} \nonumber \\
 && + \, 4 \, S_{1} \, S_{2} \, S_{-2, 1} + \frac{20}{3} \, S_{3} \, S_{-2, 1} + 
 10 \, S_{-2, 1}^2 + 10 \, S_{-2} \, S_{-2, 2} - 
 6 \, S_{1}^2 \, S_{-2, 2} + 6 \, S_{2} \, S_{-2, 2} \nonumber \\[1mm]
 && + \, 6 \, S_{1} \, S_{-2, 3} + 11 \, S_{-2, 4} - 
 6 \, S_{2} \, S_{1, 3} - 4 \, S_{1} \, S_{1, 4} - 
 4 \, S_{1, 5} + 4 \, S_{1} \, S_{2, 3} + 4 \, S_{2, 4} - 
 12 \, S_{-4, 1, 1} \nonumber \\[2mm]
 && + \, 8 \, S_{1} \, S_{-3, 1, 1} - 
 2 \, S_{-3, 1, 2} - 2 \, S_{-3, 2, 1} - 
 24 \, S_{1} \, S_{-2, -2, 1} - 20 \, S_{-2} \, S_{-2, 1, 1} + 
 16 \, S_{1}^2 \, S_{-2, 1, 1}  \nonumber \\[2mm]
 && - \, 8 \, S_{2} \, S_{-2, 1, 1} + 
 16 \, S_{1} \, S_{-2, 1, 2} - 6 \, S_{-2, 1, 3} + 
 16 \, S_{1} \, S_{-2, 2, 1} + 4 \, S_{-2, 2, 2} - 
 6 \, S_{-2, 3, 1} - 4 \, S_{1} \, S_{1, 1, 3} \nonumber \\[2mm]
 && - \, 8 \, S_{1, 1, 4} + 8 \, S_{1, 3, 2} - 8 \, S_{-3, 1, 1, 1} - 
 48 \, S_{1} \, S_{-2, 1, 1, 1} - 20 \, S_{-2, 1, 1, 2} - 
 20 \, S_{-2, 1, 2, 1} - 20 \, S_{-2, 2, 1, 1} \nonumber \\[2mm]
 && + 16 \, S_{1, 1, 1, 3} + 64 \, S_{-2, 1, 1, 1, 1} \nonumber
\end{eqnarray}

\section{Conclusions}

In a double concidence limit $x_2 \rightarrow x_1, \, x_4 \rightarrow x_3$, the four-point function of stress
tensor multiplets ${\cal T}$ is reduced to an OPE (operator product expansion) ${\cal T}(x_1)
{\cal T}(x_2) \, = \, \sum_s c(s,x) \, {\cal O}^{(s)}(x_1)$, and similarly at the other end, so that one obtains (the
sum over) the square of the structure constants  $c$ with the two-point function of the exchanged operator
${\cal O}^{(s)}$. This is a scheme invariant combination from which one can read off the structure
constants if the two-point function in the middle is assumed to be normalised to one.

Not only the primary operators but also their conformal descendents ($x$-derivatives) are exchanged.
The descendents are usually put together with the primary fields in conformal blocks called "conformal partial
waves" (cpwa). We have derived explicit results for the twist two operators in the \textbf{20'} representation
of $SU(4)$: Their anomalous dimensions come out as linear combinations of harmonic sums in full agreement
with the literature \cite{osborn, klov}. What is more, also the constants multiplying the twist two cpwa
are elaborated in terms of harmonic sums.

We have stopped short of predicting the structure constants from these results because the absolute
normalisation of the cpwa is not known\footnote{We are grateful to I.~Todorov for a discussion on this point.}.
Our results for the constant terms naturally factors into two
pieces. It is tempting to associate the first of these factors with the normalisation of the cpwa and the
second with the structure constants.

The fact that the entire result is expressed in terms of harmonic sums is a clear hint at an integrable systems
explanation, c.f. \cite{factorised,groVieira}. This issue will be addressed in future work; we are confident that the normalisation question will be understood if such an interpretation is found.

The asymptotic expansion of the individual conformal integrals is given in terms of harmonic sums with
positive indices only, and products thereof with negative powers of their argument. These results should help
with the construction of an explicit expression for the unknown integrals in terms of special functions of the
polylogarithm type. Interestingly, the cpwa decomposition leads to formulae in terms of harmonic sums only,
but here the sums can have negative indices. $S_{-1}$ does not occur, and in most of the higher sums
only the outermost index can be negative (two exceptions).

In deriving the asymptotic series by expansion by regions we have met a number of structural properties,
i.e. that given numerator terms lead to expansions that can be matched on the distinct structures $Y^{n-1} \,
S(n)/n^m, \, Y^{n-1} \, S(n)/(n+1)^m$ with or without the $m=0$ case. We have used conformal symmetry to
make four-point integrals into three-point integrals. The latter are generic by inspection; one may wonder
whether any three-point integral can be written as a spin expansion in terms of harmonic sums, or whether
the examples here are somehow specific to the ${\cal N} = 4$ SYM theory.

Obviously, our work can be extended to the twist three, four, ... trajectories corresponding to powers of
the second cross ratio $u$. We expect that the coefficients will pick up a second parameter; it remains open
for the moment whether the rational number in front of each harmonic sum will simply
start to depend on the twist, whether each trajectory is completely different, or indeed if the Euler-Zagier
sums are not sufficient to express the complete expansion.

Last, on very many occasions --- but not always --- the coefficients of a set of harmonic sums related by
index permutations add up to zero in our formulae. This hints at the existence of a special basis w.r.t. which
the results would take a much simpler form.

\section*{Acknowledgements}

This work grew out of a collaboration with P.~Heslop, G.~Korchemsky, E.~Sokatchev and V.A. Smirnov.
We are deeply indebted to V.A.~Smirnov for many discussions about the method of expansion by regions.
The author is supported by the Deutsche Forschungsgemeinschaft (DFG), Sachbeihilfe ED 78/4-1 ("eigene Stelle").

\section{Appendix: The 20 channel in terms of harmonic sums}

In the four-point correlation function at two and three loops the integrals come with very specific rational
factors depending on $Y$. It is an interesting question how the expansions in terms of harmonic sums
over powers of their arguments change by including the $Y$ expansion of the rational factors, and which
type of expansion will lead to a similar result if multiplied by such factors. In the \textbf{20'} channel we obtain
a fit of the same type as for the asymptotic expansions of the individual integrals even if the factor
$1/(1-Y)-1-Y$ from the $R$ polynomial projected onto the \textbf{20'} representation is included. We
display the result as an illustration of the universality of the basis of harmonic sums we were using, but also in
the hope that the formulae may be useful in an attempt on deriving an explicit result for the correlation function
in terms
of harmonic polylogarithms and related functions. The tree-level $1/(1-Y) + 1+O(u)$ is thus followed by the
loop correction $2 \sum_{l=1}^\infty \, a^l \, f^{(l)} + O(u)$ with the twist 2 contributions

\begin{equation}
f^{(1)} \, = \, \sum_{n=2}^\infty Y^n \biggl[ \log(u) \, \left(  - \frac{1}{2 \, n} + \frac{1}{2} \, S_{1} \right)
+\frac{1}{n^2} -S_2 \biggr] \, ,
\end{equation}

\begin{equation}
f^{(2)} \, = \, \sum_{n=2}^\infty Y^n \biggl[ \log^2(u) \, C_{2;2} + \log(u) \, C_{2,3} + C_{2,4} +
\zeta(3) \, D_{2,1} \biggr]
\end{equation}
where
\begin{eqnarray}
D_{2,1} & = &  -\frac{3}{n} + 3 \, S_{1} \, , \\ \nonumber \\
C_{2,2} & = & -\frac{1}{2 \, n^2} + \frac{1}{4} \, S_{1}^2+ \frac{1}{4} \, S_{2} \, , \\ \nonumber \\
C_{2,3} & = & \frac{3}{n^3}-\frac{S_{1}}{2 \, n^2}+\frac{S_{2}}{2 \, n}
-\frac{3}{2} \, S_{1} \, S_{2}-\frac{3}{2} \, S_{3} \, , \\ \nonumber \\
C_{2,4} & = &-\frac{6}{n^4}+\frac{S_{1}}{n^3}+\frac{S_{2}}{2 \, n^2}
+\frac{S_{1} \, S_{2}}{n}-\frac{S_{3}}{2 \, n}-\frac{2 \, S_{1,2}}{n} \\
&& - \, S_{1} \, S_{1,2}+ \frac{1}{4} \, S_{2}^2+3 \, S_{1} \, S_{3}
+\frac{13}{4} \, S_{4}-\frac{5}{2} \, S_{1,3}+3 \, S_{1,1,2} \nonumber
\end{eqnarray}
and
\begin{equation}
f^{(3)} \, = \, \sum_{n=2}^\infty Y^n \biggl[ \log^3(u) \, C_{3;3} + \log^2(u) \, C_{3;4} + \log(u) \Bigl(
C_{3;5} + \zeta(3) \, D_{3;2} \Bigr) + C_{3;6} + \zeta(3) \, D_{3;3} + \zeta(5) \, D_{3;1} \biggr]
\end{equation}
with
\begin{eqnarray}
D_{3;1} & = & \frac{25}{n} - \, 25 \, S_{1} \, , \\ \nonumber \\
D_{3;2} & = & - \, \frac{6}{n^2}+3 \, S_{1}^2+3 \, S_{2} \, , \\ \nonumber \\
D_{3;3} & = & \frac{14}{n^3}-\frac{10 \, S_{1}}{n^2}-\frac{2 \, S_{1}^2}{n}
+\frac{2 \, S_{2}}{n}+\frac{2}{3} \, S_{1}^3-8 \, S_{2} \, S_{1}
-\frac{14}{3} \, S_{3} +8 \, S_{1,2} \, , \\ \nonumber
\end{eqnarray}
\begin{eqnarray}
C_{3;3} & = &-\,\frac{1}{2 \, n^3}+\frac{S_{2}}{6 \, n}+\frac{1}{12} \, S_{1}^3
+\frac{1}{12} \, S_{1} \, S_{2}+\frac{1}{6} \, S_{1,2} \, , \\ \nonumber \\
C_{3;4} & = &\frac{6}{n^4}-\frac{S_{1}}{n^3} -\frac{S_{1}^2}{4 \, n^2}
-\frac{S_{2}}{4 \, n^2}+\frac{S_{1} \, S_{2}}{2 \, n} -\frac{S_{3}}{2 \, n}-\frac{S_{1,2}}{2 \, n} \\
&& - \, S_{1}^2 \, S_{2} - \, S_{1} \, S_{3}   -\frac{3}{4} \, S_{2}^2-\frac{3}{4} \, S_{4}
-\frac{1}{2} \, S_{1,3} \, , \nonumber \\ \nonumber \\
C_{3;5} & = &-\, \frac{30}{n^5}+\frac{6 \, S_{1}}{n^4}+\frac{S_{1}^2}{2 \,n^3}
+\frac{5 \, S_{2}}{2 \, n^3} +\frac{5 \, S_{1} \, S_{2}}{2 \, n^2}+\frac{5 \, S_{3}}{2 \, n^2}
-\frac{2 \, S_{1,2}}{n^2}-\frac{3 \, S_{1} \, S_{1,2}}{2 \, n} \\
&& +\frac{3 \, S_{1}^2 \, S_{2}}{4 \, n}+\frac{S_{1} \, S_{3}}{2 \, n}-\frac{3 \, S_{2}^2}{4 \, n}
+\frac{S_{4}}{n} -\frac{2 \, S_{1,3}}{n}-\frac{1}{2} \, S_{1}^2 \, S_{1,2}
- \, S_{1} \, S_{1,3}-2 \, S_{2} \, S_{1,2} \nonumber \\
&& + \, \frac{13}{4} \, S_{1}^2 \, S_{3}+\frac{9}{4} \, S_{1} \, S_{2}^2
+\frac{29}{4} \, S_{1} \, S_{4}+\frac{17}{4} \, S_{2} \, S_{3}+6 \, S_{5}
-5 \, S_{1,4}+ \, S_{2,3}-\frac{5}{2} \, S_{1,1,3} \nonumber \\[1mm]
&&  + \, S_{1,2,2}+ 6  \, S_{1,1,1,2} \, , \nonumber \\ \nonumber \\
C_{3;6} & = &
\frac{60}{n^6}-\frac{12 \, S_{1}}{n^5}-\frac{6 \, S_{2}}{n^4}- \frac{S_{1}^2}{n^4} 
-\frac{3 \, S_{1} \, S_{2}}{n^3}-\frac{6 \, S_{3}}{n^3}+\frac{4 \, S_{1,2}}{n^3} 
-\frac{S_{1} \, S_{1,2}}{n^2} \\
&& -\frac{3 \, S_{2}^2}{2 \, n^2}+\frac{S_{1}^2 \, S_{2}}{2 \, n^2}-\frac{6 \, S_{1} \, S_{3}}{n^2}
-\frac{8 \, S_{4}}{n^2}+\frac{10 \, S_{1,3}}{n^2}+\frac{6 \, S_{2} \, S_{1,2}}{n}
+\frac{2 \, S_{1} \, S_{1,3}}{n}-\frac{5 \, S_{1} \, S_{2}^2}{2 \, n} \nonumber \\
&& -\frac{S_{1}^2 \, S_{3}}{n}-\frac{13 \, S_{1} \, S_{4}}{2 \, n}-\frac{3 \, S_{5}}{n}
+\frac{16 \, S_{1,4}}{n}+\frac{2 \, S_{2,3}}{n}-\frac{3 \, S_{1,2,2}}{n}-\frac{125}{6} \, S_{6} 
+\frac{13}{12} \, S_{2}^3 \nonumber \\
&&  + \, 3 \, S_{1} \, S_{2} \, S_{1,2}+\frac{21}{2} \, S_{2} \, S_{1,3}+2 \, S_{3} \, S_{1,2}
+\frac{1}{2} \, S_{1}^2 \, S_{1,3}+3 \, S_{1} \, S_{1,4} - \, S_{1} \, S_{2,3}
-\frac{11}{2} \, S_{2} \, S_{1,1,2} \nonumber \\
&& - \, \frac{1}{2} \, S_{1}^2 \, S_{1,1,2}- \, S_{1} \, S_{1,1,3}-3 \, S_{1} \, S_{1,2,2}
+4 \, S_{1} \, S_{1,1,1,2}-\frac{1}{4} \, S_{1}^2 \, S_{2}^2 -9 \, S_{1} \, S_{2} \, S_{3}
-\frac{41}{4} \, S_{2} \, S_{4} \nonumber \\
&& - \, \frac{7}{2} \, S_{3}^2-\frac{17}{4} \, S_{1}^2 \, S_{4}-19 \, S_{1} \, S_{5} 
+ S_{1,2}^2+24 \, S_{1,5}+6 \, S_{2,4}+2 \, S_{1,1,4}- S_{1,2,3} \nonumber \\[1mm]
&& - \, 6 \, S_{1,3,2}-3 \, S_{1,1,1,3}+  S_{1,1,2,2}-10 \, S_{1,1,1,1,2} \, . \nonumber 
\end{eqnarray}

\end{document}